\documentclass[fleqn,twoside]{article}
\usepackage{espcrc2}
\usepackage{epsfig}

\usepackage{graphicx}
\usepackage[figuresright]{rotating}

\def\beq{\begin{equation}}
\def\eeq{\end{equation}}
\def\bea{\begin{eqnarray}}
\def\eea{\end{eqnarray}}
\def\bq{\begin{quote}}
\def\eq{\end{quote}}

\newlength{\bredde}
\def\slash#1{\settowidth{\bredde}{$#1$}\ifmmode\,\raisebox{.15ex}{/}
\hspace*{-\bredde} #1\else$\,\raisebox{.15ex}{/}\hspace*{-\bredde} #1$\fi}

\def\gappeq{\mathrel{\rlap {\raise.5ex\hbox{$>$}}
{\lower.5ex\hbox{$\sim$}}}}

\def\lappeq{\mathrel{\rlap{\raise.5ex\hbox{$<$}}
{\lower.5ex\hbox{$\sim$}}}}

\def\Toprel#1\over#2{\mathrel{\mathop{#2}\limits^{#1}}}

\newcommand{\AmS}{{\protect\the\textfont2
  A\kern-.1667em\lower.5ex\hbox{M}\kern-.125emS}}

\title{\vspace{-4.0cm}
Vector and Axial-Vector Propagators in the $\epsilon$-Regime of QCD}

\author{P.H. Damgaard\address{Niels Bohr
Institute, Blegdamsvej 17, DK-2100 Copenhagen, Denmark\vspace{-0.0cm}},
P.~Hern\'andez\address{Theory Division, CERN, CH-1211 Geneva 23, Switzerland\vspace{-0.0cm}}%
\thanks{On leave from Dpto. F\'{\i}sica
Te\'orica, Univ. Valencia.}, K. Jansen\address{NIC/DESY Zeuthen,
Platanenallee 6, D-15738 Zeuthen, Germany\vspace{-0.0cm}}, M. Laine\address{Faculty 
of Physics, University of Bielefeld, D-33501 Bielefeld, Germany\vspace{-0.0cm}},
L. Lellouch\address{Centre Physique Th\'{e}orique, CNRS, Case 907, Luminy,
F-13288 Marseille, France}}

\begin{document}

\begin{abstract}
Using quenched and unquenched chiral perturbation theory we compute
vector and axial current two-point functions at finite volume and
fixed gauge field topology, in the so-called $\epsilon$-regime of
QCD. A comparison of these results with finite volume lattice
calculations allows to determine the parameters of the corresponding
chiral Lagrangians.
\end{abstract}

\maketitle
Consider
QCD in a toroidal volume $V$ with
$L = V^{1/4}$, and assume that $V$ is large with respect to the QCD
scale, $i.e.$, $F L \gg 1$, 
where $F$ is the pion decay constant. 
Then take the chiral limit of vanishing quark mass, $m \to 0$, 
so that $m_{\pi} \ll 1/L$. This is the $\epsilon$-regime of 
QCD \cite{GL,hn}.

As in an infinite volume the lightest degrees of freedom are the Goldstone
bosons
of chiral symmetry breaking, 
described by a chiral Lagrangian,
\begin{eqnarray}
 {\cal L}_{\chi PT}\! = \! {\rm Re}\,
 {\rm Tr} \left[\frac{F^2}{4}\partial_\mu U\partial_\mu U^\dagger - 
 \Sigma\,
 U M e^{i\theta/N_f}
 \right], \nonumber
\end{eqnarray}
where $\Sigma$, $\theta, N_f$ are  
the chiral condensate, the vacuum angle, 
and the number of light flavours.

In the quenched theory, the flavour singlet field $\Phi_0\sim \ln\det U$ 
does not decouple. We thus
add, to leading order \cite{BG},
$$
  \delta {\cal L}_{\chi PT}\! = \!
 {m_0^2 \over 2 N_c} \Phi^2_0 + \frac{\alpha}{2N_{c}}\partial_{\mu}
 \Phi_0 \partial_{\mu}\Phi_0
 \;.
$$
We have performed our calculations in both the
supersymmetric \cite{BG} and replica formulations \cite{DS,DDHJ}
of quenched chiral perturbation theory, and shown
that they agree. More details can be found in ref. \cite{DHJLL}.

The $\epsilon$-regime requires an exact evaluation of the zero momentum
mode integrals. We split up ($\xi(x)$ are non-zero momentum modes):
\begin{equation}
U(x) ~=~ \exp\left[\frac{2i\xi(x)}{F}\right] U_0 \;,
\end{equation}
and perform the $U_0$ integration exactly.

At the quark level the vector and axial vector currents are
\begin{eqnarray}
V^a_\mu(x) &\equiv& \bar{\psi}(x) i \gamma_\mu T_{N_v}^a  \psi(x) \;, \cr
A^a_\mu(x) &\equiv& \bar{\psi}(x) i \gamma_\mu\gamma_5 T_{N_v}^a \psi(x) ~,
\end{eqnarray}
and we add sources to extract them from the effective theory,
\begin{eqnarray}
\partial_\mu U \! \rightarrow \! \partial_\mu U + 
i (v^a_\mu - a^a_\mu) T_{N_v}^a U - i U T_{N_v}^a (v^a_\mu + a^a_\mu).
\nonumber
\end{eqnarray}
Correlation functions are computed by taking derivatives of the partition
function w.r.t. these sources in the chiral effective theory.
We have determined the relevant 2-point functions up to and including
next-to-leading order in the 
$\epsilon$-expansion, in the sense of refs.~\cite{GL,DDHJ}.
The gauge field topological charge is fixed to be $\nu$ by
Fourier transformation w.r.t. $\theta$ of the partition function.

While we present results here for the vector and axial currents 
correlators only, similar computations are well motivated and
have been carried out for other observables as well, such as 3-point
functions containing weak operators~\cite{ml,pm}. 
Various numerical techniques for the $\epsilon$-regime have 
recently been discussed in~\cite{ml}. 
Preliminary measurements of correlation functions in quenched
QCD have already been reported~\cite{PO,CJN,NCJ}.

\bigskip
\noindent
{\bf Predictions for full QCD:}

It turns out that all zero-mode integrals
can be expressed in terms of the finite-volume chiral
condensate
\begin{eqnarray}
\Sigma_{\nu}(\mu) ~=~ \frac{\Sigma}{N_f}\langle
{\mbox{\rm Re Tr}}[U_0]\rangle_{\nu,U_0} \;,
\end{eqnarray}
and derivatives thereof. Here $\mu \equiv m\Sigma V$ and we also
define, to one loop,
\begin{eqnarray}
\mu' ~\equiv~ \mu\left(1 + \frac{N_f^2-1}{N_f}\frac{\beta_1}
{F^2\sqrt{V}}\right)\;,
\end{eqnarray}
where $\beta_1$ is a shape-dependent (but universal) constant
\cite{GL,HL}. We get
\begin{eqnarray}
\int\! {\rm d}^3 \vec{x} \,
\langle {V}^a_0(x) {V}^a_0(0) \rangle_\nu &=& \cr
&& \hspace{-3.8cm} -\frac{F^2}{2 T}\left\{{\cal J}_- +
\frac{N_f}{F^2} \left(\frac{\beta_1}{\sqrt{V}} {\cal J}_- 
-\frac{T^2}{V} k_{00} {\cal J}_+ \right) \right\} , 
\label{eq:cvunq}\\
\int\! {\rm d}^3 \vec{x} \,
\langle {A}^a_0(x) {A}^a_0(0) \rangle_\nu 
& = & \cr
&& \hspace{-3.8cm}-\frac{F^2}{2 T}\left\{{\cal J}_+ +
\frac{N_f}{F^2} \left(\frac{\beta_1}{\sqrt{V}} {\cal J}_+ 
-\frac{T^2}{V} k_{00} {\cal J}_- \right) \right.\nonumber\\
&& \hspace{-3.8cm}+ \left. \frac{4 \mu}{N_f F^2} \frac{T^2}{V}h_1(\tau) 
\langle {\rm Re}\,{\rm Tr} [U_0]\rangle_{\nu,U_0}  \right\},
\label{eq:caunq}
\end{eqnarray}
where 
$h_1(\tau) \equiv [(|\tau|-{1}/{2})^2
-{1}/{12}]/2$,
$\tau \equiv x_0/T $, 
$T$ is the extent of the lattice in the time direction, and $k_{00}$
is a numerical factor \cite{HL,H}. Analogous results but
without projecting onto fixed topological charge $\nu$ were
first computed by Hansen \cite{H}. The expectation values 
\begin{eqnarray}
{\cal J}_{+} & = & \frac{1}{N_f^2-1} \left( N_f^2 - 2 + 
\langle {\rm Tr}[U_0] {\rm Tr}[U_0^\dagger]\rangle_{\nu,U_0} \right)\nonumber\\
{\cal J}_{-} & = & \frac{1}{N_f^2-1} \left( N_f^2 + 
\langle {\rm Tr}[U_0] {\rm Tr}[U_0^\dagger]\rangle_{\nu,U_0} \right)
\end{eqnarray}
are known analytically \cite{DDHJ},
\begin{eqnarray}
\langle {\rm Tr}[U_0] {\rm Tr}[U_0^\dagger]\rangle_{\nu,U_0} &=& N_f 
\left[ \frac{\Sigma_\nu'(\mu')}{\Sigma}\right. \cr
&& \hspace{-3cm}\left.+ N_f \left(\frac{\Sigma_\nu(\mu')}
{\Sigma}\right)^2
+\frac{1}{\mu'} \frac{\Sigma_\nu(\mu')}{\Sigma} - 
\frac{\nu^2 N_f}{\mu'^2} \right],
\end{eqnarray}
where $\Sigma_{\nu}(\mu)$ can be expressed explicitly in terms
of the modified Bessel functions $I_n(x)$. 

\bigskip
\noindent
{\bf Predictions for quenched QCD:}

We now consider $N_v$ valence quarks embedded in a theory of $N_f$ quarks
in total, and then take the replica limit $N_f \to 0$ \cite{DS,DDHJ} 
(results agree with what one obtains by the supersymmetric formulation 
\cite{BG}). Remarkably,
all contributions from the famous double-pole propagator of
quenched $\chi$PT {\em cancel} exactly at both leading and
next-to-leading order. We thus only need the usual massless
pion propagator,
\begin{equation}
{\bar \Delta}(x) \equiv \frac{1}{V} \sum_{p\neq 0} 
\frac{e^{i p x}}{p^2} \;.
\end{equation}

Let 
${\cal O}^{a,-}_\mu(x) \equiv  V^{a}_\mu(x)$ and 
${\cal O}^{a,+}_\mu(x) \equiv A^{a}_\mu(x)$, 
and define $t^a_\pm \equiv T_{N_v}^a \pm U_0 T_{N_v}^a U_0^{-1}$.
Then
\begin{eqnarray}
\left\langle {\cal O}^{a,\sigma}_\mu(x)~ 
 {\cal O}^{b,\tau}_\zeta(0) \right\rangle_\nu 
 & = &  \cr
&&\hspace{-4cm} - \frac{F^2}{2} 
 \sigma\tau \left\langle {\rm Tr} [t^a_{\sigma} t^b_{\tau}]
 \right\rangle_{\nu,U_0} 
 \; {\partial_{\mu}}{\partial_{\zeta}} 
 \bar{\Delta}(x)  \nonumber\\
 & & \hspace*{-4cm} - \frac{m\Sigma}{4} \sigma\tau
 \left\langle {\rm Tr}[\{ t^a_{\sigma}, t^b_{\tau}\}(U_0+ U^{-1}_0)] 
 \right\rangle_{\nu,U_0} 
 \cr
&& \hspace{-4cm}\times   
 \int {\rm d}^4 z~\partial_{\mu} \bar{\Delta}(z-x) 
 \partial_{\zeta} \bar{\Delta}(z) ~.
\end{eqnarray}
As in full QCD, the required zero-mode integrals are known in closed
analytical form, and the result can be expressed in terms of \cite{DOTV}
\begin{eqnarray}
{\Sigma_\nu (\mu) \over \Sigma} \!\equiv\! 
\mu \Bigl[ I_\nu (\mu) K_\nu (\mu) + I_{\nu +1} (\mu)
K_{\nu -1}(\mu) \Bigr] +{\nu \over \mu}, \nonumber
\end{eqnarray}
where also $K_n(x)$ is a modified
Bessel function. As will become clear below, we 
will not need $\mu'$ and the one-loop corrected condensate \cite{D01}
(with its problematical finite volume logarithm) in the quenched case,
since the correction is of one order higher than what we are computing.

\begin{figure}[t]
\epsfxsize=7.cm\epsffile{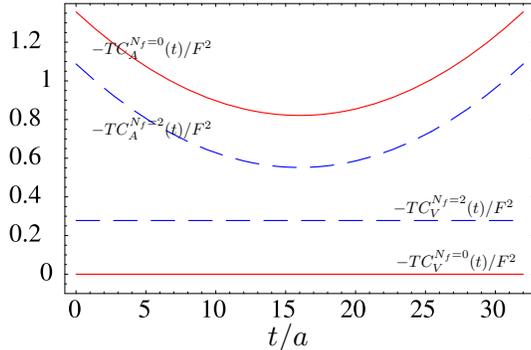}
\vspace{-1.cm}
\caption{\em Quenched and $N_f=2$ 
correlation functions
of Eqs.\ \ref{eq:cvunq} and \ref{eq:caunq}
and Eqs.\ \ref{eq:cvunq} and 
\ref{eq:caunq} for $\nu=0$ plotted vs 
$t/a$. Here
$a=2\,\mathrm{GeV}$, $L=24a$, $T=32a$, 
$m=0.1\,\mathrm{MeV}$, $F=93\,\mathrm{MeV}$
and $\Sigma=(250\,\mathrm{MeV})^3$.}
\vspace{-0.5cm}
\label{fig:cacv}
\end{figure}

Including next-to-leading order, and making use of the exact
results for the quenched zero-momentum mode integrals of 
ref.~\cite{TV}, we find
\begin{eqnarray}
\left\langle{V}^a_0(x)~ {V}^a_0(0) \right\rangle_\nu 
 & = & 0,  \\
 \int\!{\rm d}^3 \vec{x} \,
 \left\langle 
 {A}^a_0(x)~ {A}^a_0(0) \right\rangle_\nu 
 & = & - \frac{F^2}{T}\biggl[ 1 + \cr
&& \hspace{-3.5cm} \frac{2\, m \Sigma_\nu(\mu) T^2}{F^2} 
 h_1\left(\frac{x_0}{T}\right) \biggr] ~.
\end{eqnarray}
The vector-vector correlator of quenched QCD thus vanishes identically
up to and including next-to-leading order. Examples of these correlation
functions are plotted in Figure \ref{fig:cacv} along with their unquenched
counterparts of Eqs.\ \ref{eq:cvunq} and \ref{eq:caunq}.

{\em Measuring the axial-vector--axial-vector correlation in this 
$\epsilon$-regime of QCD directly gives the pion decay constant $F$
to leading order. At next-to-leading order also the infinite-volume
chiral condensate $\Sigma$ can be extracted.}

\bigskip
\noindent
{\bf An argument to all orders:}

The quenched $\langle V^a(x)V^b(y)\rangle$ correlation function must
vanish to all orders. This follows from the following argument.
There are two ways to contract external quark lines to generate
the 2-point functions. One is the
``connected'' contraction, where the quarks flow from point
$x$ to point $y$. The other is the ``disconnected'' contraction,
where the quarks flow back to the starting points $x$ or $y$. In the
quenched approximation no other quark flow topologies are possible.

Now consider the singlet correlator in the full theory, 
in the replica limit $N_f \to 0$. It is then easy to  
see that the disconnected piece vanishes in this limit. 
Therefore the quenched {\em flavor non-singlet} correlator, 
which also only gets contributions from the connected piece,
can, up to an overall factor, be computed in the {\em singlet
sector} of the full theory. But the singlet vector current 
vanishes identically because the corresponding source does not couple to
the pion field, and thus the non-singlet quenched vector
correlator is also zero.

\end{document}